\documentclass[useAMS,usenatbib,fleqn]{mn2e}

\usepackage{graphicx}
\usepackage{amsmath,amsfonts,amssymb}
\usepackage{color}
\usepackage{epsfig}

\usepackage{hhline}
\usepackage{array,booktabs}
\usepackage{epstopdf}
\usepackage[T1]{fontenc}

\newcommand{\msun}{M$_{\rm \odot}$}

\newcommand{\apj}{ApJ}
\newcommand{\apjl}{ApJL}
\newcommand{\mnras}{MNRAS}
\newcommand{\aj}{AJ}
\newcommand{\prd}{Phys.\ Rev.\ D.}
\newcommand{\prl}{Phys.\ Rev.\ Lett.}
\newcommand{\cqg}{Class.\ Quantum Grav.}

\title[Black Hole Binaries formed in Clusters]{Black Hole Binaries Dynamically Formed in Globular Clusters}
\author[Park et al.]{\parbox{\textwidth}{Dawoo Park$^{1}$\thanks{Email: dawoo@astro.snu.ac.kr}, Chunglee Kim$^{2}$\thanks{Email: chungleekim@kasi.re.kr}, Hyung Mok Lee$^{1,3}$\thanks{Email: hmlee@astro.snu.ac.kr}, Yeong-Bok Bae$^{2}$\thanks{Email: baeyb@kasi.re.kr}, and Krzysztof Belczynski$^{4}$\thanks{Email: chrisbelczynski@gmail.com}}\\
$^{1}$Astronomy Program, Department of Physics and Astronomy, Seoul National University, 1 Gwanak-ro, Gwanak-gu, Seoul 08826, Korea\\
$^{2}$Korea Astronomy \& Space Science Institute, 776 Daedeokdae-ro, Yuseong-gu, Daejeon 34055, Korea\\
$^{3}$Centre for Theoretical Physics, Seoul National University, 1 Gwanak-ro, Gwanak-gu, Seoul 08826, Korea\\
$^{4}$Astronomical Observatory, University of Warsaw, Aleje Ujazdowskie 4, 00-478 Warsaw, Poland}

\begin{document}
\date{Draft version, 2017 March}
\pagerange{\pageref{firstpage}--\pageref{lastpage}} \pubyear{2017}
\maketitle
\label{firstpage}
\begin{abstract} 
We investigate properties of black hole (BH) binaries formed in globular clusters via dynamical processes, using {\it direct} N-body simulations. We pay attention to effects of BH mass function on the total mass and mass ratio distributions of BH binaries ejected from clusters. Firstly, we consider BH populations with two different masses in order to learn basic differences from models with single-mass BHs only. Secondly, we consider continuous BH mass functions adapted from recent studies on massive star evolution in a low metallicity environment, where globular clusters are formed. In this work, we consider only binaries that are formed by three-body processes and ignore stellar evolution and primordial binaries for simplicity. Our results imply that most BH binary mergers take place after they get ejected from the cluster. Also, mass ratios of dynamically formed binaries should be close to one or likely to be less than 2:1. Since the binary formation efficiency is larger for higher-mass BHs, it is likely that a BH mass function sampled by gravitational-wave observations would be weighed toward higher masses than the mass function of single BHs for a dynamically formed population. Applying conservative assumptions regarding globular cluster populations such as small BH mass fraction and no primordial binaries, the merger rate of BH binaries originated from globular clusters is estimated to be at least 6.5 yr$^{-1}$ Gpc$^{-3}$. Actual rate can be up to more than several times of our conservative estimate. 
\end{abstract}
\begin{keywords}
{gravitational waves -- stars: black holes -- galaxies: star clustesr: general -- methods: numerical}
\end{keywords}

%%%%%%%%%%%%%%%%%%%%%%%%%%%%%%%%%%%%%%%%%%%%%%%%%%%%%%%%%%%%%%%%%%
% Introduction
%%%%%%%%%%%%%%%%%%%%%%%%%%%%%%%%%%%%%%%%%%%%%%%%%%%%%%%%%%%%%%%%%%
\section{Introduction}
During the first observing run of the Advanced LIGO detector (from September 12, 2015 to January 19, 2016), two black hole binary (BBH) mergers at cosmological distances were discovered. The individual (source-frame) black hole (BH) masses of the first event, labeled as GW150914, are m$_1 \simeq 36.2$ \msun, m$_2 \simeq 29.1$ \msun~\citep{Abbott16a}. The second event (GW151226) consists of relatively less massive BHs of m$_1 \simeq 14.2$ \msun\,and m$_2 \simeq 7.5$ \msun~\citep{Abbott16b}. The estimated masses of final BHs formed by merging of two BHs are $62.3$ \msun\ (GW150914) and $20.8$ \msun\ (GW151226), respectively. Discoveries of GW150914 and GW151226 proved the existence of BHs in compact binaries. Furthermore, GW150914 provides with observational evidences for BHs with masses between $30-70$ \msun\, which is significantly larger than the estimated BH masses in Galactic X-ray binaries (e.g., Tanaka 2000\nocite{Tanaka00}). With two confirmed detections and one BBH candidate (LVT151012), however, the observations have not provided constraints for the origins of known BBHs \citep{Abbott16c} or BH mass distribution \citep{Abbott16d}.

Many authors have attempted to explain the formation and evolution of BBHs based on a standard massive binary evolution scenario (e.g., \citealt{bel10,mink15,bel16a}). After the detection of GW150914, different ideas were suggested such as chemically homogeneous evolutionary channels for BBHs \citep{mink16}, primordial BHs as progenitors of GW150914 \citep{sasaki16} and the roles of Population II BBHs \citep{inayoshi16}. 

Alternatively, dynamical processes in dense stellar systems such as globular clusters (GCs) have been suggested (e.g., \citealt{ban10,mor13,tan13,Bae14}). Recently, \citet{rodriguez2016a,rodriguez2016b} performed Monte Carlo simulations for GCs and studied evolution of BBHs and BH properties. They obtained BBH merger rate of $\sim5$ Gpc$^{-3}$ yr $^{-1}$ in the local Universe, mentioning that their estimated rate can be decreased if significant natal kicks are exerted on BH populations in the cluster. One of the interesting implications of their work is that the majority of BBHs would have mass ranges between $32-66$ \msun, and the contribution of cluster-origin BBHs would be roughly 14 per cent among the all {\it merging} BBHs in the local Universe. Adapting similar Monte-Carlo models, \citet{chatterjee2017} examined how different initial assumptions affect the cluster's evolution and properties of BH populations. They concluded that initial mass function, stellar wind, natal kicks are most important factors to affect the number of BBHs or properties of BH populations formed in clusters. 

Recent work by \citet{hurley2016} examined properties of BHs and neutron star (NS) populations by using a {\tt NBODY6} code. In particular, the authors took into account natal kicks for BHs between 0 and 100 km~s$^{-1}$. They found that the cluster properties such as number density of stars and relaxation time-scale \citep{spi87} affect the properties of BBHs formed in the cluster significantly. In addition, \citet{banerjee2016} examined BHs in dense stellar clusters applying the latest N-body simulation code ({\tt NBODY7}) with post-Newtonian corrections in orbit calculations. Taken into account different cluster masses, metallicities, stellar wind and evolution scenaria, and half-mass radius sizes corresponding to young massive and starburst clusters, the authors concluded that many BBH mergers could take place within the cluster.

One of the simplifications made by many N-body simulations is the assumption of a fixed-mass for all BHs: typical choice is 10 \msun~motivated from the average mass of observed BHs from Galacic X-ray binaries\footnote{https://stellarcollapse.org/bhmasses/} However, population synthesis works predict that a BH mass distribution can range about $3 - 100$ \msun. Similar to discs of galaxies, GCs are also expected to contain BHs with a wide range of masses. There are attempts to include stellar evolution and/or BH mass distribution in N-body simulations. For example, \citet{tan13} examined the effects of BH mass distribution in cluster dynamics, utilizing a prescription for binary evolution included in {\tt NBODY4} with modifications for mass losses of massive stars at supernova explosions following \citet{eldridge04}. The BH mass function used in \citet{tan13} spans from 5 to 20 \msun\ with a peak at about 4 \msun, which is relatively narrower than what was suggested for disc populations.

This work is a follow-up of Bae et al.\ (2014, Paper I hereafter)\nocite{Bae14}. Paper I investigated properties of NS-NS and BBHs formed in and ejected from clusters by N-body simulations. The main conclusions of Paper I are: (a) significant fraction of BHs are ejected from a cluster after core collapse and about 30 per cent of ejected BHs are in binaries, (b) the merger rate of ejected BBHs per a cluster is $2.5 - 10$ Gyr$^{-1}$. This corresponds to a detection rate of $15 - 60$ yr$^{-1}$ for the advanced LIGO-Virgo network. (c) NS-BH binary is not likely to be formed in a cluster. In this work, we focus on BBHs formed in GCs, considering two-component and continuous BH mass functions.

The organization of this work is as follows. In \S \ref{section:bbh}, we describe the evolution of BHs in a cluster qualtitatively. In \S \ref{section:model}, we present our models and assumptions. The results of our simulations are summarized in \S \ref{section:results}. Lastly, we discuss implications of results in \S \ref{section:discussion}.

%%%%%%%%%%%%%%%%%%%%%%%%%%%%%%%%%%%%%%%%%%%%%%%%%%%%%%%%%%%%%%%%%%
% BBHs                        
%%%%%%%%%%%%%%%%%%%%%%%%%%%%%%%%%%%%%%%%%%%%%%%%%%%%%%%%%%%%%%%%%%
\section{Formation and Evolution of Black Hole Binaries in Clusters} \label{section:bbh}
BHs are formed by the evolution of stars more massive than ~20 $M_\odot$ (e.g., Mirabel 2017\nocite{Mirabel17}) whose lifetime is expected to be shorter than $10^7$ yrs. Considering the fact that most of the Milky Way (MW) GCs are older than 10 Gyr, the formation of BHs took place during very early phase of the evolution of a cluster. As soon as BHs are formed, therefore, they become most massive components within a cluster that is still young. BHs segregate into the cluster core due to dynamical friction then form a dense BH subsystems in time-scale rather quickly (e.g.\ Lee 1995, 2001a\nocite{Lee95,Lee2001a}; Breen \& Heggie 2013\nocite{bh2013}). The BH subsystem in the central part then undergoes core collapse rapidly, further increasing the core density composed of BHs. Frequent close encounters among BHs are expected to take place in the core, including formation of BBHs and their interactions with surrounding stars.

In order to form binaries, orbital energy of two nearby stars has to be taken away. This can happen either via dissipative or non-dissipative processes. For extended stars such as main-sequence stars or giants the orbital energy can be transformed into internal energy via tidal interactions (e.g., \citealt{pt77,lo1986}). For the case of compact stars such as NSs and BHs, gravitational waves (GWs) emitted during close encounters can be only possibility of dissipating orbital energy (e.g., Quinlan \& Shapiro 1989\nocite{Quinlan1989}). The non-dissipative process for the formation of binaries requires a temporary triple system which eventually becomes a bound system of two stars by ejecting the third star at high velocity. This is called binary formation by a three-body process.

The direct capture rate per unit volume between two identical stars with mass $m$ and number density $n$ can be expressed as
\begin{equation}
{\dot n}_{\rm cap}={dn_{\rm cap}\over dt}= {1\over 2} n^2 \left< \Sigma_{\rm cap} v_{\rm rel} \right>~,
\end{equation}
where $\Sigma_{\rm cap}$ is the capture cross section, $v_{\rm rel}$ is a relative velocity of two approaching stars, and the brackets represent an average taken over a given velocity distribution. The gravitational radiation capture cross section has been obtained for using post-Newtonian approximation following \citet{Quinlan1987}. For two identical stars,
\begin{equation}
\Sigma_{cap} \approx 17{G^2 m^2\over c^{10/7} v_{\rm rel}^{18/7}}~,
\end{equation}
where $c$ is the speed of light. The binary formation rate via three-body processes has been obtained by \citet{Goodman93} as follows,
\begin{equation}
	{\dot n}_{3b}={dn_{3b}\over dt} \approx C n^3 {(Gm)^5 \over\sigma^9}~,
	\label{eq:3bd}
\end{equation}
where $\sigma$ is the one-dimensional velocity dispersion of the system and $C\approx 0.75$.

Assuming that $\left<v_{\rm rel}^{2}\right> = 2 \sigma^{2}$, the ratio between gravitational radiation capture and binaryformation by three-body processes can be obtained from eqs.\ (1) and (3).
\begin{equation} 
	{{\dot n}_{\rm cap}\over {\dot n}_{3b}} \approx 0.38 \left({10^5 {\rm~pc}^{-3} \over n}\right) \left({10 {\rm M}_\odot\over m}\right)^3\left({\sigma\over 10 {\rm~km~s^{-1}}}\right)^{52/7}~.
\end{equation}
Now, clusters with $\sigma$ greater than 10 ${\rm km~s}^{-1}$ are rare. Even for clusters with $\sigma$ greater than 10 ${\rm km~s}^{-1}$, the velocity dispersion of BHs will be smaller than that of stars whose masses are much smaller than BHs. Of course, BHs that are less massive than 10 \msun\ could form binaries via gravitational radiation. Also the gravitational radiation capture could be efficient, if the stellar density is much lower than $10^5 {\rm pc^{-3}}$. We note that, however, binary formation rate in a low density environment would be low regardless of the nature of processes available in a cluster. In short, we expect that most BBHs formed by dynamical processes would have gone through (multiple) three-body interactions in typical GC conditions.

Next we consider the orbital evolution of binaries formed by three-body interactions. Once a binary is formed, it can interact with surrounding stars rather frequently since the effective cross section becomes larger for binaries. Close encounters between a single star and a binary are defined as those with the pericenter approach of an order of the semi-major axis of a binary. For each encounter, let's assume some fraction of binding energy of a binary is transformed into translational energies of participating members, i.e.,
\begin{equation}
	(\Delta E )_{\rm enc} = \xi | E_B| = \xi{Gm^2\over 2 a}~,
\end{equation}
where $a$ is the semi-major axis of the binary and $\xi$ is the fraction of energy extracted from a binary. Through these processes a binary becomes tighter and tighter if the binary is `hard', i.e.\ $v_{\rm orb} >> \sigma$, where $v_{\rm orb}$ and $\sigma$ are the binary's orbital velocity and velocity dispersion of surrounding stars, respectively. In contrast, if a BBH is dynamically soft, the binary would be resolved into two single BHs via interactions with another BH. Since the absolute value of the binding energy becomes larger during hardening process, the binary as well as the single star gain more translational energy and eventually get ejected from the cluster, if the binary's recoil velocity becomes larger than the cluster's escape velocity $v_{\rm esc}$. Typically, the mean velocity of ejected BBHs is about 1.8 times of $v_{\rm esc}$ (see fig.\ 8 in Paper I for a velocity distribution of ejected binaries at the tidal radius normalised by $v_{\rm esc}$). Assuming that the net linear momentum relative to the cluster's center of mass motion remains to be zero during and after the binary-single interaction and $\xi=0.4$ \citep{Lee2001b}, the condition for a binary's critical semi-major axis at ejection can be written as follows: 

%\begin{equation}
\begin{eqnarray}
a<a_{\rm crit} &\approx& {Gm\over 15 v_{\rm esc}^2} \nonumber \\
               &\approx& 2.2 \times 10^{13} {\rm cm}  \left({m\over 1~{\rm M}_\odot}\right) \left( 20 ~{\rm km~s}^{-1}\over v_{\rm esc}^2\right)^2~.
\end{eqnarray}\label{acrit}
We can see that the binary's orbit is quite tight when it is ejected.

The orbit of an initially tight binary further decreases (slowly) by the emission of GWs. Whether the binary will get ejected or merge within the cluster is determined by the comparison between typical time-scale for binary-single encounter and the GW induced merger. For typical physical parameters for GCs, the binary-single encounter time-scale is shorter than GW induced merger time-scale\citep{Lee2001a}. Thus we may assume that the BBHs in GCs get ejected from the cluster with typical semi-major axis given by eq. (\ref{acrit}) and undergo passive evolution afterward outside of the cluster by emitting GWs. In the following sections, we confirm that BBHs formed in GCs indeed follow the evolutionary path as outlined above.

%%%%%%%%%%%%%%%%%%%%%%%%%%%%%%%%%%%%%%%%%%%%%%%%%%%%%%%%%%%%%%%%%%
% Model
%%%%%%%%%%%%%%%%%%%%%%%%%%%%%%%%%%%%%%%%%%%%%%%%%%%%%%%%%%%%%%%%%%
\section{N-body simulation} \label{section:model}

For N-body simulations, we utilize {\tt{NBODY6}} code \citep{aar03} with graphical processing units (GPUs) on NVIDIA Tesla C2075 platforms or NVIDIA Tesla C1060 platforms. N-body simulation is computation-intensive. One of the main limitations of N-body approach is the maximum number of stars that can be realized. The number of stars in a MW GC is typically ${10^{6}-10^{7}}$. N-body simulations are often based on a simplified cluster model with a smaller number of particles, $N \sim {\it O}$($10^{4-5}$) up to millions presented by recent work \citep{wang2016}. Clearly the adopted number of stars in our study is smaller than that of real clusters by almost two orders of magnitude. Our choice of smaller N is mainly due to the limitation of the currently available computing resources. It is well known that the dynamical evolution driven by the two-body relaxation is independent on N when time scale is expressed in units of the half-mass relaxation time as shown in eq.\ (\ref{eq:trh}) (see for example, Cohn 1980\nocite{Cohn1980}). Therefore we expect that our results on global evolution would not be affected by the choice of small N. However, the evolution binary orbits would not be proportional to the two-body relaxation time scale. Instead, the collision time between binaries and singles would be more relevant. This issue will be discussed in \S 4.1.

The purpose of our work is not to perform N-body simulations with realistic modeling of GCs. Instead, we attempt to obtain statistically robust results on the characteristics of BBHs as a function of global properties of clusters. In particular, we pay special attention to hardness and eccentricity of ejected binaries. The hardness is the binding energy of binaries in units of (3/2) $m \sigma^2$, which is an average kinetic energy of surrounding stars (see eq.\ (7) in \S 4.1 for the definition of hardness). As we showed in the previous section, the critical value of the semi-major axis for the ejection (a$_{\rm crit}$) is given as Gm/$v_{\rm esc}^{2}$, (see eq.\ (6)). Therefore the hardness of an ejected binary corresponding to a$_{\rm crit}$ is proportional to $v_{\rm esc}^{2} /{\sigma}^2$. This does not depend on N, and only weakly depends on the shape of cluster's density distribution. When we estimate the a$_{\rm crit}$ in the Section 3, we made a very simplified assumption that 40 per cent of binding energy is released per encounter. In reality, the fractional change of binding energy varies significantly depending on the details of each encounter. Our simulations provide more realistic distribution of semi-major axes of the ejected binaries that can be converted into the distribution of hardness. The eccentricity distribution is also known as independent of the details of the simulations. Conversion of hardness to semi-major axis can be done for a given cluster by adopting the observed value of the velocity dispersion \citep{gnedin02}. We still need a large number of stars in simulations, in order to avoid statistical fluctuations. We repeat $\sim$ 10 simulations per model with $N=2.5\times 10^{4}$ or $5\times 10^{4}$, so that we can reduce statistical fluctuations significantly. For a given model, all parameters are fixed and only random seed numbers are different per each run.

\subsection{Initial conditions for model clusters}
We use similar conditions used in Paper I for controlling initial and environmental conditions in our simulations (see section 2 in Paper I for details). We use a King model for an initial density profile of a cluster \citep{king66} with a fixed concentration parameter $W_{0} = 6$. The cluster is laid on a static external tidal field created from a point mass galaxy at 8.5 kpc distance and 220 km~s$^{-1}$ circular velocity. 

As discussed in \S \ref{section:bbh}, we only consider three-body processes in this work. Stellar evolution and primordial binaries are not included; Paper I showed that cluster dynamics with BHs, except for time-scale of the evolution itself, are not sensitive to different assumptions on the concentration parameter or primordial binary fractions (see their figs.\ 4 and 5). Moreover, some binary properties are insensitive to the particle number of N-body simulations. Examples of such properties include eccentricity, hardness, and velocity distributions of ejected binaries (figs. 6-8 in Paper I). We can use those number independent parameters to investigate the dynamically formed binary populations. We use a reference parameter set used in Paper I, except using different BH mass functions.

%%%%%%%%%%%%%%%%%%%%%%%%%%% table1
\begin{table}
\begin{center}
	\caption{Model parameters for a two-component BH mass functions assumed in this work. The first column lists model names. Individual BH masses ($m_1$ and $m_2$), ratios between total masses ($M_1 : M_2$), and total BH mass fractions in the cluster are shown in the second through fourth columns.}
\label{table:model1}
\begin{tabular}{@{}llcccc}
\toprule
Model & $m_{1}$, $m_{2}$ & $M_{1}$: $M_{2}$ & BH mass fraction \\ 
name  & (\msun)          &                  & (\%)       \\
\toprule
A1    & 10, 20           & 2:1   & 2.5  \\	
A2    &                  & 5:1   & 2.5  \\
A3    &                  & 2:1   & 5.0  \\		
A4    &                  & 5:1   & 5.0  \\
A5    &                  & 2:1   & 1.35 \\	
A6    &                  & 5:1   & 1.35 \\
\\
B1    &  5, 10           & 2:1   & 2.5  \\	
B2    &                  & 5:1   & 2.5  \\
B3    &                  & 2:1   & 5.0  \\	
B4    &                  & 5:1   & 5.0  \\
\bottomrule
\end{tabular}
\end{center}
\end{table}
%%%%%%%%%%%%%%%%%%%%%%%%% table1(end)

\subsection{BH mass function}
The cluster is assumed to be composed of normal stars represented by the ones with 0.7 \msun~and black holes whose masses are much higher than normal stars. We employ (a) simple mass function composed of only two mass components, and (b) continuous mass spectrum spanning from 5 to $\sim$40 \msun~for the black hole population.

The two-component BH mass function is the simplest expansion from the Paper I where all BHs are assumed to be 10 \msun. We set $m_{1} < m_{2}$ for individual BH masses of $m_1$ and $m_2$. Summary of model parameters for two-component BH mass functions is given in Table \ref{table:model1}. We consider two cases: (a) models A1-A6 consist of ($m_{1},m_{2}$)$=$(10, 20) \msun~BHs in the cluster and (b) models B1-B4 consist of ($m_{1},m_{2}$)$=$(5, 10) \msun\, BHs. Models with odd numbers (A1, A3, and A5 for instance) are assumed to have a total mass ratio of $M_{1}(\equiv n_1 m_1) : M_{2}(\equiv n_2 m_2) = 2:1$. For even numbered models, we consider $M_{1} : M_{2} = 5:1$ (A2, A4, A6, B2 and B4), where $M_{\rm i}$ is the total sum of all BHs with mass $m_{\rm i}$ and number $n_{i}$ ($i=1, 2$). We consider A1 and A2 as references.

Parameters of continuous BH mass functions are listed in a Table~\ref{table:model2}. 
The continuous BH mass distribution is based on recent studies of high mass stars with different metallicity. BH mass is calculated from a set of hydrodynamical supernova models; these models were used to form a simple formulae for rapid evolutionary calculations \citep{fryer2012}. BH mass depends on {\em (i)} initial star mass (e.g., Kroupa \& Weidner 2003)\nocite{kw03}, {\rm (ii)} wind and eruptive mass loss (as observed in Luminous Blue Variables) during stellar evolution (e.g., Vink et al.\nocite{vink2001}; Humphreys \& Davidson 1994\nocite{hd1994}) and {\rm (iii)} mass loss during final core collapse and potential supernova explosion \citep{fryer2012}. We adopt a rapid supernova explosion model that can successfully reproduce the mass gap observed between NSs and BHs (the lack of compact objects in $2-5$ \msun\ range; Belczynski et al.\ 2012)\nocite{bel12}. This model allows for BH formation either through partial fall back (weak supernova and some mass ejection) or direct collapse of a star to a BH (no mass is ejected during BH formation). The minimum BH mass depends on the supernova engine model. We adopt a model with the minimum BH mass of 5 \msun~in this work. For low metallicity stars the maximum BH mass of a stellar-origin depends mostly on wind mass loss and the extend of initial mass function (150 \msun\ adopted here). The wind mass loss depends sensitively on metallicity \citep{bel10}. The BH maximum mass may be limited by pair-instability pulsation supernova mass loss to about $\sim 40-50$ \msun\ for Population I/II stars (e.g., Belczynski 2016b)\nocite{bel16b}. Models adopted in this work produce the maximum BH mass of 41.5 \msun\ for two metallicities ($Z = 0.1 Z_{\rm \odot}$ and $0.01 Z_{\rm \odot}$). The mass fractions used in N-body simulations are set close to 5 per cent for the continuous mass function (see the third column of the Table~\ref{table:model2}). Because the mass function used in the N-body simulation has been realized by Monte Carlo sampling, the actual values used in each simulation are not precisely 5 per cent, though.

%%%%%%%%%%%%%%%%%%%%%%%%% table2 (begin)
\begin{table}
    \caption{Parameters of continuous BH mass function models. The first column lists model name. The second and third columns present the particle number and BH mass fraction in each model. The ranges of BH masses in each distribution are listed in the fourth column. Metallicity assumed in each distribution and the number of N-body runs are listed in the fifth and sixth columns.}
	\label{table:model2}
	\begin{tabular}{@{}llcccccc}
		\toprule
		Model & $N_{\rm total}$ & BH & BH mass & Assumed &$N_{\rm run}$ \\
		 & $(\times {\rm 1,000})$ & fraction(\%) & range & [Fe/H] &  \\
		\toprule
		Belc01 & 50 & 4.4 $\sim$ 5.1\  & $5 - 41.5$ & $-1$ & 8 \\
		Belc001 & 50 & 4.6 $\sim$ 5.7\ & $5 - 41.5$ & $-2$ & 8 \\
		\bottomrule
	\end{tabular}
\end{table}
%%%%%%%%%%%%%%%%%%%%%%%%% table2 (end)

%%%%%%%%%%%%%%%%%%%%%%%%%% fig1(begin)
\begin{figure*}
\centering
\includegraphics[width = 1.7\columnwidth]{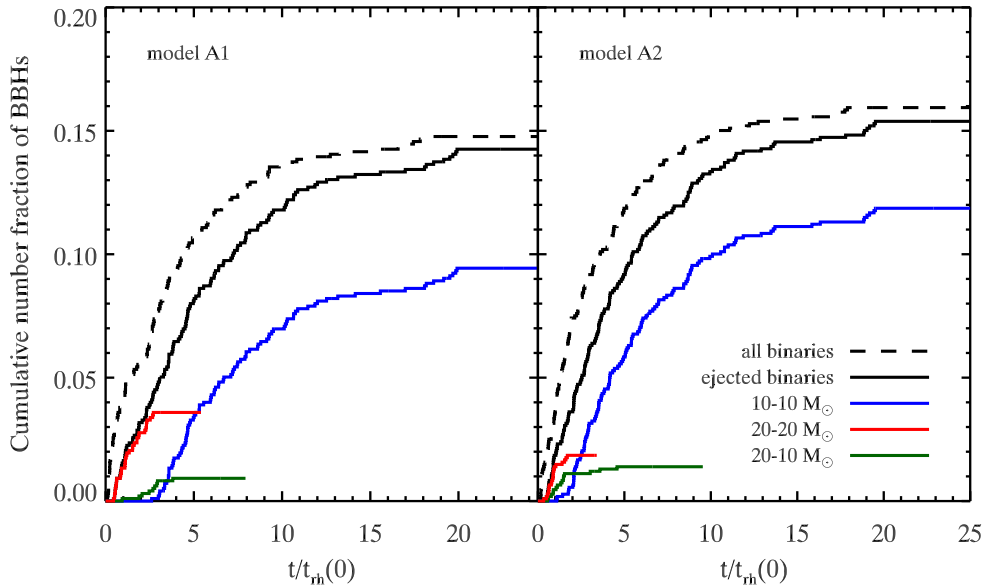}
\caption{Cumulative number of BBHs obtained from two-component BH mass functions (models A1 and A2). Black dashed lines stand for the cumulative number of all BBHs formed in a cluster. Black solid lines represent the cumulative number of {\it ejected} BBHs among those formed in the cluster. For comparison, we also show results for ejected BBHs with three different masses in colour: $10-10$ \msun\ (blue), $20-20$ \msun\ (red), and $20-10$ \msun\ (green) binaries.}
\label{fig:formej}
\end{figure*}
%%%%%%%%%%%%%%%%%%%%%%%%%% fig1(end)

%%%%%%%%%%%%%%%%%%%%%%%%%%%%%%%%%%%%%%%%%%%%%%%%%%%%%%%%%%%%%%%%%%
% Results                        
%%%%%%%%%%%%%%%%%%%%%%%%%%%%%%%%%%%%%%%%%%%%%%%%%%%%%%%%%%%%%%%%%%
\section{Results}\label{section:results}

Binaries are formed at the center of a GC as the density becomes high by dynamical friction and subsequent core collapse. First, most massive components in the cluster form binaries. Then, ejection of binaries begins to take place right after the formation of the first binary. It means that hardening of binary orbits is a rapid process (more rapid than the binary formation time-scale). Not only binaries but also single BHs are ejected from the cluster since the recoil energy during a binary-single encounter is shared by the binaries and singles. If all BH masses in a cluster are the same, the recoil velocity of a single BH is expected to be twice of that of a binary. This implies that the number of single BHs ejected from a cluster is larger than that of binaries. Indeed, Paper I found that only 30 per cent of the ejected BHs is belonged to binaries while the rest (70 per cent) are single BHs. 

Except for quantitative details, lower-mass BHs follow similar paths that we described earlier for higher-mass BHs such as mass segregation, binary-single (i.e., three-body) interaction, and ejection from the cluster after most of the higher-mass BH get ejected in the form of either binaries or singles. In addition to those discussed in the Paper I such as cumulative numbers of ejected binaries, hardness distribution, and the ratio of merging binaries with respect to the all ejected binaries (figs. 4, 7, 9 in the Paper I), we examine number and mass ratios between BBHs with different masses in this work.

\subsection{Two-component BH mass functions}

%about fig1
In Fig.\ \ref{fig:formej}, we compare cumulative number distributions of different BBH populations as a function of time (in units of the cluster's initial relaxation time-scale $t_{\rm rh}(0)$). Here we show results obtained from models A1 (the left panel) and A2 (the right panel), where $M_{1}:M_{2}=2:1$ and $5:1$, respectively. Both models consist of $M_{1}=10$ \msun\ and $M_{2}=20$ \msun\ BHs. The black dashed lines are all BBHs formed in the cluster and black solid lines are those ejected. Colour lines represent cumulative numbers of BHs in binaries with different masses. The formation and ejection of BBHs occur hierarchically with respect to their masses. The highest-mass binaries ($20-20$ \msun, red curves) are formed first at the early phase of the cluster's evolution (less than a few $t_{\rm rh}(0)$). When the highest-mass BBH populations are nearly completely ejected from the cluster, the next massive binaries ($20-10$ \msun) are formed and ejected (green curves). Lastly, the lowest-mass BBHs ($10-10$ \msun) are formed and ejected (blue curves). We note that unequal-mass BBHs can be formed but rare. Fig.\ \ref{fig:formej} implies that, by the time 10 \msun~BHs start to interact with, there are only few 20 \msun\ BHs are left in the central part of the cluster.

Note that there is nearly a constant time gap between all BBHs (black dashed lines) and ejected binaries (black solid lines). This time gap can be interpreted as a typical lifetime of a BBH spent in the cluster before ejection. If we wait long, almost all BBHs are ejected from the cluster and the black solid and dashed lines asymptotically converge (with one or two binaries left in the cluster). However, our results show that the total number of binaries formed in the cluster is slightly larger by a few than that of ejected. This is presumably attributed to dissolved binaries as BBHs are formed and dissolved through multiple three-body interactions. The fraction of dissolved binaries comprises less than 3 per cent of the total BBHs formed in a cluster.

We emphasize a few things from Fig.\ \ref{fig:formej}. Firstly, we confirm unequal BBHs can be formed, rarely though, and ejected from a cluster. Secondly, if the total mass ratio between the two BH populations are larger (i.e., model Bs with $M_{1}:M_{2}=5:1$), more unequal-mass BBHs can be produced because of the shorter time gap between mass segregations of $m_{1}$ and $m_{2}$. Thirdly, there is {\it always} some delay between the formation (dashed) and ejection (black solid) of BBHs from a cluster. Lastly, almost all BHs are eventually ejected from the cluster ($t > 20 t_{\rm rh}$). Results from model B's are similar to Fig.\ \ref{fig:formej}.

\begin{figure}
\includegraphics[width = \columnwidth]{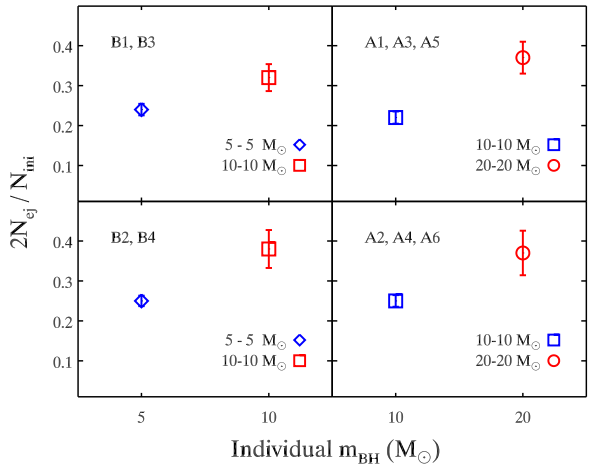} 
\caption{Number fraction of BHs in ejected binaries normalised by the initial number of BHs in the cluster. The x-axis is the individual BH masses ($m_1$ and $m_2$) in \msun. We show simulation results with error bars. We show results from models assuming two-component BH mass functions (upper panels with $m_1:m_2=1:2$ and lower panels with $m_1:m_2=1:5$). BH masses are $m_{1,2}=(5, 10)$ \msun~(left panels from model Bs) or $(10,20)$ \msun~(right panels from model As).}
\label{fig:frac1}
\end{figure}

%about fig2
Fig.\ \ref{fig:frac1} shows the number of BHs in ejected binaries ($2N_{\rm ej,i}$) normalised by the initial BH number $N_{\rm ini,i}$ for each mass $i=1,2$. For example, in the top left panel, we compare $2N_{\rm ej,5 M_{\odot}}/(N_{\rm ini,5 M_{\odot}})$ (blue diamond) and $2N_{\rm ej,10 M_{\odot}}/(N_{\rm ini,10 M_{\odot}})$ (red square) obtained from models B1 and B3. Here, we show equal-mass binaries only. Higher-mass BBHs ($m_2$-$m_2$), shown as red squares (left panels) and red circles (right panels), are more preferentially formed and ejected than lower-mass binaries (blue diamonds in left panels and blue squares in right panels). As shown in eq.\ (\ref{eq:3bd}), the BBH formation is sensitive to the mass of individual components (${{\dot n}_{\rm 3b}}\propto m^{5}$). By comparing $2N_{\rm ej,i}/N_{\rm ini,i}$ for each mass, the number fraction of BHs in binaries with respect to the initial number is {\it larger} for a more massive component $m_2$.

%%%%%%%%%%%%%%%%%%%%%%%%%% fig2(begin)
%%%%%%%%%%%%%%%%%%%%%%%%%% fig2(end)

\begin{figure}
\includegraphics[width = \columnwidth]{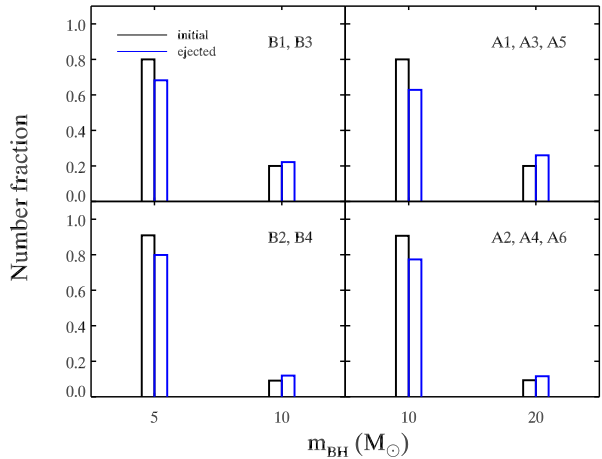}
\caption{Number fraction of BHs ($m_1$ and $m_2$, respectively) of the initial and ejected population, i.e., $N_{\rm ini,i}/N_{\rm ini}$ (black bars) and $2N_{\rm ej, i}/N_{\rm ej}$ (blue bars) where $i=1,2$. We consider the same set of models presented in Fig.\ \ref{fig:frac1}. Notice that the heights of black and blue bars (i.e.\ number fractions) between $m_1$ and $m_2$ are reversed. See text for details.}
\label{fig:nbhs}
\end{figure}

%about fig3
In Fig.\ \ref{fig:nbhs}, we compare the number of initial BHs normalised by the total initial number of BHs for each mass (black bars), i.e., $N_{\rm ini,m_{i}}/N_{\rm ini}$ where $i=1,2$. The blue bars represent the number of ejected BHs (either in binaries or in singles) normalised by the total ejected BHs for each mass, i.e., $N_{\rm ej,m_{i}}/N_{\rm ej}$, where $N_{\rm ej}=N_{\rm ej,m_1} + N_{\rm ej,m_2}$. As expected from Fig.\ \ref{fig:frac1}, the fraction of BHs in binaries among the ejected is larger than that of initial populations, if the BH mass is higher ($m_2$). For example, we obtain $N_{\rm ini}(m_{1}): N_{\rm ini}(m_{2})=4:1$ from model B. For example, we have $N_{\rm ini}(m_{1}): N_{\rm ini}(m_{2})=0.8:0.2$ for models B1, B3 (the top left panel), but $N_{\rm ej}(m_{1}): N_{\rm ej}(m_{2}) \sim 3.5:1$ for ejected BBHs from the same models ($N_{\rm ej}(m_{1}): N_{\rm ej}(m_{2})\simeq0.7:0.2$). This means that the mass function derived from the observed binaries would be significantly biased toward the higher-mass components when compared to the intrinsic mass function of single BHs within the observable volume for GW detectors.

\begin{figure}
\includegraphics[width=\columnwidth, height=6.8cm]{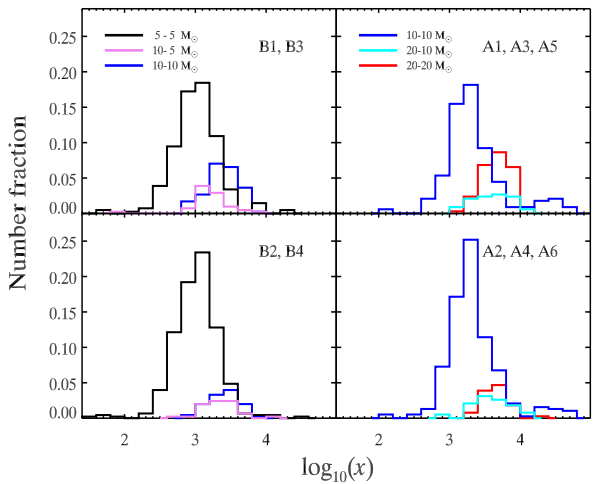}
\caption{Hardness distributions of ejected BBHs for different models with the two-component BH mass function.}
\label{fig:hardness}
\end{figure}

%about fig4
In Fig.\ \ref{fig:hardness}, we present distributions of dimensionless hardness of ejected BBHs from different models. Hardness of a binary represents a compactness of a binary and is defined by masses of binary components, binary separation ($a$) and velocity dispersion of cluster stars ($\sigma$), and the mass of ordinary stars ($m_{\rm *}=0.7$ \msun) as shown in the equation below. \begin{equation}x=\frac{G m_{1} m_{2}/2a}{3 m_{\rm *} \sigma^{2}/2}~.\end{equation}\label{eq:hardness} Note that the clusters would have $v_{esc}^2 \sim 12 \left<\sigma^2\right>$ (see for example, Spitzer 1987\nocite{spi87}). If we use the condition for the separation $a$ in eq.\ (6), hardnesses of ejected binaries would be $x>60 \left(m_B/m_*\right)$ where $m_B$ is the BH mass (i.e., $m_1$ or $m_2$. If $m_1 = m_2=10$ \msun\ BHs and $m_*=0.7$ \msun, a $10-10$ \msun\ binary's hardness $x$ would be about 840. The histogram shown in Fig.\ \ref{fig:hardness} is consistent with our expectations. For a given binary population, its hardness distribution follows nearly Gaussian with $\log_{10}(x)$. Assuming two-component BH mass function, our results are still consistent with Paper I that is assumed a single BH mass of 10 \msun. In general, the averaged hardness obtained from ejected BBHs approximately proportional to the binary's total mass. As expected, locations of peaks in hardness distributions are different for each population (colours indicate different masses as presented in Fig.\ 4). We find that, for a two-component BH mass function, BBHs' hardness distribution is likely to be scalable by mass by comparing top (bottom) and left and top (bottom) right panels in Fig.\ \ref{fig:hardness}. 

%%%%%%%%%%%%%%%%%%%%%%%%%% fig3(begin)
%%%%%%%%%%%%%%%%%%%%%%%%%% fig3(end)

Binaries with large hardnesses (i.e. BBHs in tight orbits) are good candidates for GW detections. In particular, some of the ejected BBHs are expected to merge within a Hubble time. These {\it merging} binaries can be GW sources for advanced LIGO-type detectors. Binaries formed and ejected from a cluster can have orbital eccentricities following thermal distribution between zero to one (e.g.\ figure 6 in Paper I). For each ejected BBH, we convert its dimensionless separation obtained from N-body simulations to the binary separation in a physical unit utilizing a cluster's central velocity $\sigma(0)$. Incorporating a binary's eccentricity and separation at the time of ejection from the cluster to eqs.\ (8-9), we compute the binary's merging time-scale \citep{peters1964}. Then we determine a merging binary, if its estimated merging time-scale is shorter than a Hubble time.
\begin{equation} \label{eq:tmerge} t_{mrg}=\frac{5}{64}\frac{c^5 a^4}{G^3 m_{1}m_{2}(m_{1}+m_{2})f(e)}~, 
\end{equation}
where $f(e)$ is defined as follows
\begin{equation} f(e)=(1-e^2)^{-7/2}\left({1+\frac{73}{24}e^2+\frac{37}{96}e^4}\right)~.  
\end{equation}

%%%%%%%%%%%%%%%%%%%%%%%%%% fig4(begin)
%%%%%%%%%%%%%%%%%%%%%%%%%% fig4(end)

\begin{figure}
\includegraphics[width=\columnwidth, height=6.8cm]{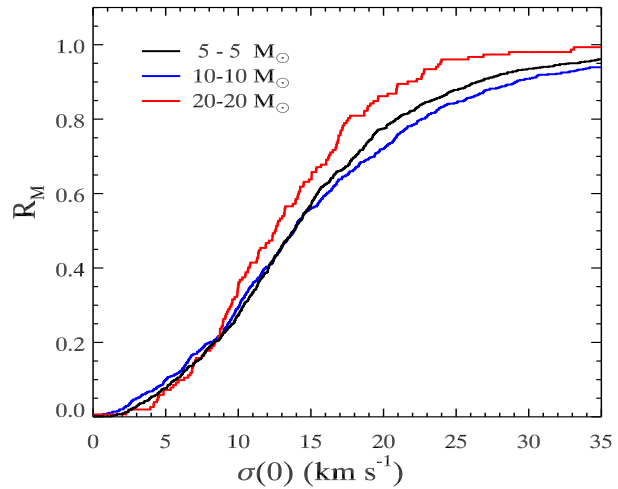}
\caption{Ratio of merging BBHs to the total ejected binaries as a function of the central velocity dispersion of a GC.}
\label{fig:merge}
\end{figure}

%about fig5
Fig.\ \ref{fig:merge} presents the fraction of binaries that can merge within a Hubble time among the ejected binaries, $R_{\rm M}$, as a function of cluster's central velocity dispersion. We show results for three equal-mass BBH populations: $5-5$ \msun\ (black), $10-10$ \msun\ (blue), $20-20$ \msun\ (red), respectively, obtained from all models listed in Table 1. The fraction of merging binaries $R_{\rm M}$ increases along with $\sigma(0)$ and almost converges to one around $\sigma(0) \sim 35$ km~s$^{-1}$. We redirect readers to fig. 9 in the Paper I for central velocities of the observed GCs in the Milky Way and results of $R_{\rm M}$ for models with 10 \msun~BHs only.

Let's examine detectability of BBHs ejected from clusters considering GW detectors on the Earth. Following Paper I, we compute the source event rate (a.k.a.\ merger rate) using the following eq.\ (\ref{eq:r_gc}) below.
\begin{equation}\label{eq:r_gc} {\cal R}_{\rm GC}=\sum_{i=1}^{n} \frac{M_{\rm i}}{\langle m_{\rm *}\rangle} f_{\rm n} \sum_{j=1}^{3} f_{\rm eb, j} R_{\rm M, j}(\sigma_{i}(0)) /(n_{\rm GC} t_{\rm GC})~, 
\end{equation}
where $M_{\rm i}$ is the total mass of each cluster $i$ adopted from a catalog containing $n_{\rm GC}=141$ clusters in the Milky Way \citep{gnedin02}, $\left<m_{\rm *}\right>$ is the mean mass of cluster stars, $f_{\rm n}$ is the number fraction of BHs initially introduced in the cluster, $f_{\rm e,b,j}$ is a number fraction of a BBH population, where $j$ stands for three BBH populations: $m_1-m_1$ (j=1), $m_2-m_2,$ (j=2) and $m_1-m_2$ (j=3). For equal-mass binaries (j=1,2), we read $R_{\rm M, j}(\sigma_{i}(0))$ from Fig.\ 5 given $\sigma_{i}(0))$ for each MW GC. For unequal binaries ($j=3$ such as $5-10$ \msun) we take an average value of $R_{\rm M, j=1}$ and $R_{\rm M, j=2}$ from Fig.\ 5. Then, $\frac{M_{\rm i}}{\langle m_{\rm *}\rangle} f_{\rm n} \sum_{j=1}^{3} f_{\rm eb, j} R_{\rm M, j}(\sigma_{i}(0))$ represents the total number of merging binaries ejected from the $i$-th cluster. Lifetime of all MW GCs are assumed to be $t_{\rm GC}=13$.

The expected number of binary merger events for different combination of BH masses depends on many factors. As we have described, BBHs with higher-mass components are likely to form more efficiently than lower mass populations. The merger fraction estimated from ejected binaries can slightly affect the merger rate. Since the discrepancies between $R_{\rm M}$ curves are not significantly large, we may expect that the GW sources will be more weighted toward the higher masses. In addition, the horizon distances for higher-mass BBHs are larger by approximately a factor of binary's mass ratio, we may expect to detect higher-mass binaries more frequently than simple expectation from a BH mass function. In that sense, surprisingly large BH masses estimated from GW150914 would not be a simple coincidence.

%%%%%%%%%%%%%%%%%%%%%%%%%% fig5 (begin)
%%%%%%%%%%%%%%%%%%%%%%%%%% fig5 (end)

\begin{figure}
\vspace*{-.3cm}
\includegraphics[width=\columnwidth, height=7.5cm]{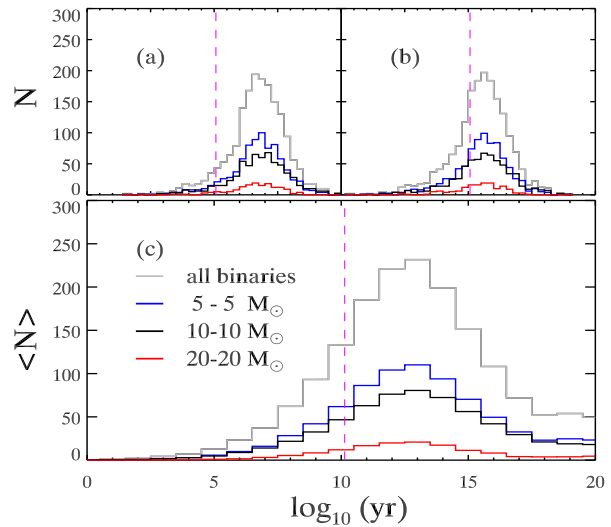}
\caption{Distribution of delay time, which is the time span between the ejection and the epoch of merge for each BBH, calculated for a given central velocity dispersion $\sigma(0)$ of the cluster. Panels (a) and (b) present delay time distributions assuming $\sigma(0)$ = 5 and 10 km~s$^{-1}$, respectively. Panel (c) shows a delay time distribution for ejected, merging BBHs from all models, averaged by the observed $\sigma(0)$ of MW GCs.The pink dashed vertical line indicates a Hubble time (13.7 Gyr).}
\label{fig:delay}
\end{figure}

In Fig.\ \ref{fig:delay}, we plot histogram of delay times for ejected, merging BBHs. The delay time of a binary can be defined by the duration between the epoch of ejection of a binary from a cluster and the merger event. In this work, a binary's delay time is essentially the merger time-scale we described earlier. We show results for specific values of $\sigma(0)=5$ and 10 km~s$^{-1}$ in the upper panel, respectively. The shape of the delay time distribution is nearly independent of $\sigma(0)$. The mean delay time decreases for a cluster with larger central velocity. As shown in Fig.\ \ref{fig:merge}, only some fraction of ejected binaries are to merge within a Hubble time for a typical velocity dispersion observed from MW GCs. In Fig.\ \ref{fig:delay}, the peak is located at $t >$ Hubble time. Among the merging binaries, at least a few per cent of ejected binaries are expected to merge within $10^{7}$ yrs, much quicker than others. We also point out that, although it is not impossible, only very few binaries would merge within a cluster. We may safely assume that majority of mergers takes place {\it outside} of the cluster.

Considering results obtained from models listed in Table 1, based on two-component BH mass functions, we calculate BBH merger rate ${\cal R}_{\rm GC}$ (per GC per Gyr) and rate volume density ${\cal R}_{\rm V}$ (per Gpc$^3$ per yr), where ${\cal R}_{\rm V}=\rho_{\rm GC}{\cal R}_{\rm GC}$, $\rho_{\rm GC}$ is the specific number density of GCs per galaxy. Following Paper I, we assume  $f_{\rm n} = 0.001$ and adopt $\rho_{\rm GC}=8.4 h^3 {\rm Mpc}^{-3}$ \citep{por00}. For all models considered in this work, we obtain ${\cal R}_{\rm GC}\sim2.5$ Gyr$^{-1}$ per GC. This is similar to reference results from Paper I based on the similar set of parameters. Recall that we ignore primordial binaries and assume the BH mass fraction of $1.35-5.0$ in this work. We find that assumptions on mass in a two-component BH mass function within a factor two, i.e., (5,10) \msun\ versus (10,20) \msun, does not affect to rate estimates. The corresponding rate volume density from our result is ${\cal R}_{\rm V}\sim6.5$ ${\rm Gpc}^{-3}{\rm yr}^{-1}$.

We note that the estimated merger rate in this work is likely to be a lower limit because of our very conservative assumptions. The GC population we see today in the Milky Way could be only a small fraction of initial cluster populations since many clusters could have been disrupted by various reasons (e.g., Lee \& Ostriker 1995, Gnedin \& Ostriker 1997)\nocite{LeeGoodman1995,go1997}. These disrupted GC population could have produced significant number of BBHs as comparible as those from observed clusters. Furthermore, many GCs are formed in low metallicity environments where stellar mass function could have been more weighted toward higher-mass stars, i.e., top-heavy mass function suggested by Marks et al.\ 2012)\nocite{marks12}. In addition, the existence of primordial (tight) binaries can also increase BBH merger rate. Our main assumptions are similar to the reference model in Paper I. We note that Paper I argued that the actual merger rate of BBHs formed and ejected from clusters can be up to four times higher than their rate estimates obtained from the reference model.

%%%%%%%%%%%%%%%%%%%%%%%%%% fig6(begin)
%%%%%%%%%%%%%%%%%%%%%%%%%% fig6(end)

\subsection{Continuous BH mass function}\label{section:cont}
Let us emphasize two important lessons from the two-component models for BHs in GCs. One is that the dynamical process favours the formation of equal-mass binaries and the other is that the formation efficiency for higher-mass BBHs is larger than lower-mass populations. In reality, however, a continuous mass distribution for BHs would be expected in a cluster. The main uncertainty is the range and shape of the distribution. As an expansion of our study, we investigate the dynamical evolution of BHs formed in GCs, assuming a continuous BH mass function generated by population synthesis.

\begin{figure}
\includegraphics[width=\columnwidth, height=6.8cm]{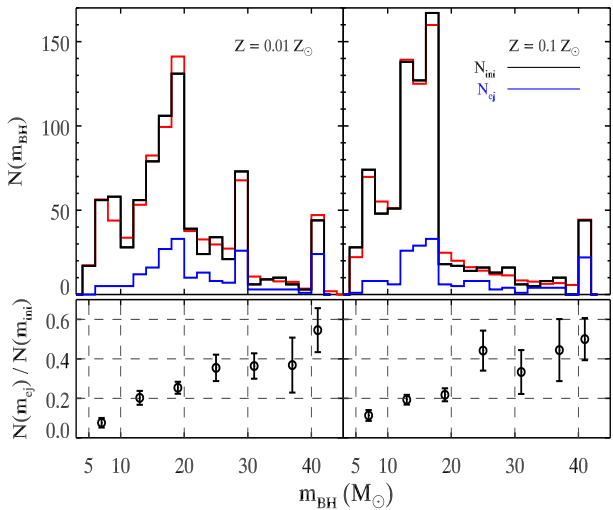}
\caption{Upper panels : BH mass functions Belc001 ($Z=0.01~Z_{\rm \odot}$, left) and Belc01 ($Z=0.1~Z_{\rm \odot}$, right). Black solid lines refer to the initial BH mass function that are randomly sampled from original results obtained by population synthesis (red lines). Blue solid lines stand for histograms of BH masses obtained from ejected binaries. All histograms are plotted with a 2-\msun\ bin. Lower panels : Ratio between the number of each mass bin read from the ejected (blue) and initial (black) population with error bars $\sqrt(N(m_{\rm ej})/N(m_{\rm ini})$. The ratios are plotted with a 6-\msun\ bin for simplicity.}
\label{fig:inej}
\end{figure}

The BH mass functions obtained from ejected binaries (blue histograms in Fig.\ \ref{fig:inej}) are multimodal, and most prominent peaks are found at around $10-20$ \msun. For both metallicities ($Z = 0.1 Z_{\rm \odot}$ and $0.01 Z_{\rm \odot}$), the initial BH mass ranges are between 5 and 41.5 \msun. Efficient formation of higher-mass BBHs is apparent from lower panels of Fig.\ \ref{fig:inej}. In our simulations, the BH formation rate for 40 \msun\ BH is about 3 times higher than that of 10 \msun\ BBHs for both metallicity cases. This means that the mass function derived from GW data alone would be highly skewed toward a upper end, compared to the actual BH mass function.

\begin{figure}
\vspace{-.6cm}
\includegraphics[width=\columnwidth]{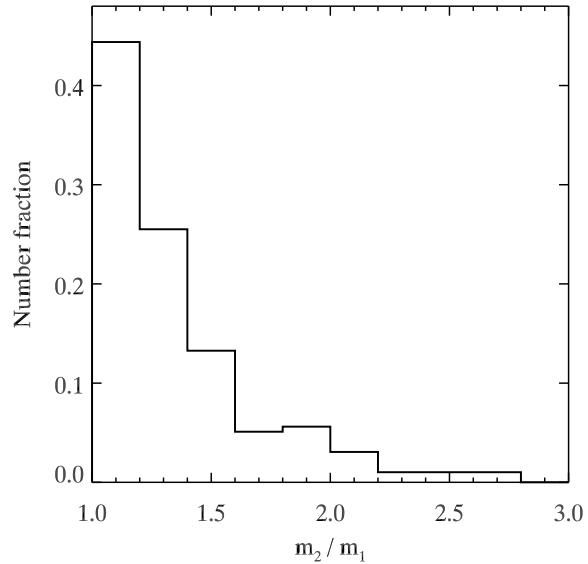}
\caption{Number of ejected, merging BBHs as a function of the mass ratio $m_{2}/m_{1}$ ($m_1 < m_2$). The results are obtained from continuous BH mass functions (Belc01 and Belc001) as described in the text. As expected, equal-mass BBHs are dominant. Some BBHs do have mass ratios between $\sim2-3$.}

\label{fig:cont}
\end{figure}

Similar to two-component mass models, formation of nearly equal-mass BBHs are strongly preferred by continuous BH mass functions (see Fig.\ \ref{fig:cont}). Our results show that most BBHs formed and ejected from a cluster have a mass ratio $q=m_{2}/m_{1} (m_{1}<m_{2})$ less than 2. Among those ejected and merging BBHs, only 6 per cent of them have $q>2$. Now, the mass ratio of known BBH is 1.2 (GW150914) and 1.9 (GW151226), respectively. Since estimation of individual masses from GW observation has substantial errors, we cannot say whether these two observations are consistent with dynamical binaries based on the mass ratio only. However, as the number of detected sources increase the statistical errors would decrease.

%%%%%%%%%%%%%%%%%%%%%%%%%%%%%%%%%%%%%%%%%%%%%%%%%%%%%%%%%%%%%%%%%%
% Discussion
%%%%%%%%%%%%%%%%%%%%%%%%%%%%%%%%%%%%%%%%%%%%%%%%%%%%%%%%%%%%%%%%%%
\section{Discussion} \label{section:discussion}
Compact binaries are important sources of GWs. In a dense stellar environment like GCs, compact binaries consisting of NSs and BHs can be formed by dynamical processes. In this work, we expand the scope from Paper I and examine the evolution of BBHs with various BH mass distributions in a GC.

For all cluster models, we have assumed that BHs and ordinary stars follow the same form of initial density profiles. In reality, density profiles of more massive components (i.e.\ BHs) could have been more centrally concentrated than lower mass components. However, the difference between the evolutions of initially mass-segregated clusters and un-segregated clusters is likely to be negligible, as the time-scale for BH mass segregation would be rather short. In general, the evolution of GCs can be described as follows. The most massive component in the system rapidly loses its energy and sinks toward the central part of the cluster through dynamical friction. This mass segregation time-scale is shorter than the cluster's relaxation time by a factor of $m_{\rm h}/\left<m\right>$, where $m_{\rm h}$ and $\left<m\right>$ are the highest component's mass and the mean mass of all stars, respectively (see for example, Lee 1995, 2001a\nocite{Lee95,Lee2001a}). In our models, BHs are more massive than ordinary stars by large factors. After mass segregation, the central part of a cluster is dominated by the highest-mass population. Close interactions among highest-mass components lead to the formation of binaries.

%%%%%%%%%%%%%%%%%%%%%%%%%% fig7(begin)
%%%%%%%%%%%%%%%%%%%%%%%%%% fig7(end)

%%%%%%%%%%%%%%%%%%%%%%%%%% fig8(begin)
%%%%%%%%%%%%%%%%%%%%%%%%%% fig8(end)

Typically, the time in N-body simulations is normalised by the cluster's initial half-mass relaxation time $t_{\rm rh}$(0). For instance, the time $t$ in Fig.\ 1 is normalised by $t_{\rm rh}$(0). Following \citet{spi87}, the half-mass relaxation time of a cluster is defined as
\begin{equation}
t_{\rm rh} = 0.138 \frac{N^{1/2} r_{\rm h}^{3/2}}{G^{1/2} m_{\star}^{1/2} ln(\gamma N)}~,
\label{eq:trh}
\end{equation}
where $r_{\rm h}$ is the half-mass radius of a cluster. $N$ is the total number of stars. $m_{\star}$ is the average mass of all stars in the cluster and $\gamma=0.11$ is adapted from \citet{gie1994} as described in Paper I.

We assume that the variation of $t_{\rm rh}$ is very mild at least in the context of our simulations. In the early phase of a cluster, $t_{\rm rh}$ increases slowly because of the expansion of the cluster due to the heating effect of the BBH. However, as the cluster loses mass $t_{\rm rh}$ decreases slowly. Overall, we may assume that $t_{\rm rh}$ remains nearly constant within a factor of two. Known GCs in the Milky Way have diverse values of $t_{\rm rh}$, ranging from $10^8$ to more than $10^{10}$ yrs, with a median value being around $1.35\times10^{9}$ yrs. This means that the time-scale for complete depletion of BHs in GCs also varies from cluster to cluster. 

Our simulation is attempted to establish GCs at present based on observations. BH depletion could have occured earlier in the GC's evolution. The binary-single encounters that lead to the hardening and ejection of binaries take place in dynamical time. The relaxation time of a GC is proportional to $N\times t_{\rm dyn}$, where $N$ is the number of stars and $t_{\rm dyn}$ is the dynamical time-scale of the cluster. Since our $N$ used in simulations is much smaller than actual number of stars in clusters, the difference between relaxation time and dynamical time is smaller than actual cluster environments. In our simulations, the binaries are ejected after the formation in about $2\sim4$ relaxation time-scale. If the number of stars were larger than our simulations, the time gap between binary formation and ejection would have been shorter than our results (as shown in Fig.\ 1). Our simulations indicate that the complete ejection of BHs is about 20 $t_{\rm rh}$(0) but the actual ejection could require shorter amount of time. We note that the $\sim20$ $t_{\rm rh}$(0) is the expected lifetime of clusters in steady tidal field (e.g., Henon 1961\nocite{henon1961}, Lee \& Ostriker 1987\nocite{lo1987}). The cluster's lifetime can be shortened significantly by the presence of stellar mass function \citep{LeeGoodman1995} or tidal shocks (e.g., Gnedin, Lee, \& Ostriker 1999)\nocite{glo1999}. The initial rotation of the clusters also accelerate the time to core collapse by factors of several \citep{kls2004}, as well as the cluster's lifetime \citep{kim2008}. In view of such possibilities, we may assume that the clusters expected lifetime would be 10 $t_{\rm rh}$(0) or short.

As a corollary, we also expect that BHs would have been completely ejected from a cluster if its $t_{\rm rh}$ is relatively short (i.e., $< 10^{9}$ yrs). Clusters with sufficiently long $t_{\rm rh}$ could still contain a substantial number of BHs either in the forms of binaries or singles. Considering the simplifications made in this work, it is not impossible that some clusters can harbour BHs that are more massive than the maximum BH mass expected from mass star evolution. That is, in some clusters, BBHs are formed in a very early phase of the cluster. If the BBHs consist of the highest-mass BHs in the cluster and if the binary is hard enough, the binaries can have chances to merger via emitting GWs within the cluster unless the recoil velocity during the merger is larger than $v_{\rm esc}$.

It is known that the majority of massive stars including BH progenitors are in binaries (e.g., \citealt{sana12}). In Paper I, the authors varies the number of ejected BBHs from a cluster varying the fraction of primordial binaries from zero per cent up to 50 per cent. They found no significant differences in results (see figure 5 in the Paper I). Based on their results in the Paper I, the authors concluded that BBH properties (examined by N-body simulations) are not sensitive to primordial binary fraction. \citet{chatterjee2017} also found outcomes from Monte-Carlo simulations have barely changed by different primordial binary fractions. In the present work, we consider a conservative assumption that all clusters have single BHs initially. 

Our N-body simulations take into account only Newtonian prescriptions of binary orbits. Therefore, gravitational radiation capture (two-body process) is not included in this work, hence, all binaries are formed by three-body processes. If post-Newtonian corrections are taken into account in N-body simulation, there could be rare cases of binaries formed by two-body processes similar to what described in \citep{HongLee2015}. These binaries are likely to be formed in the central region of a galaxy, where stellar number density is significantly higher and stars can have larger velocity dispersions than those in GCs. If a GC would have high enough central number density, some BBHs can be formed by two-body interactions capture as well as three-body interactions. Binaries formed by two-body processes are particularly interesting GW sources because their orbits could be extremely eccentric: typically $e \sim  1- 6\times 10^{-6} \left({\sigma\over 10 {\rm km/s}}\right)^{10/7}$ at the time of formation. The eccentricities decrease rapidly during the inspiral phase, but significant amount of eccentricity would remain when the gravitational wave frequency enters the detectors sensitive band. Such a possibility was discussed in detail by O'Leary et al.\ (2009)\nocite{oleary09} and Hong \& Lee (2015)\nocite{HongLee2015} in the context of galactic nuclei star clusters. Observation of eccentric merger from ground based detectors would suggest the presence of the captured binaries, although we expect that they would be very rare.

GW observations of BBHs will provide information on underlyaing properties such as spins and masses. We ignore the effects of kicks in this work. For heavy BHs considered in this work, natal kicks from supernova explosions are most likely to be small or negligeable as predicted by theoretical works (e.g. Fyer et al.\ 2012)\nocite{fryer2012}. However, based on EM or GW observations, low-to-moderate natal kicks ($\leq 300$ km s$^{-1}$) even for massive BHs can not yet be excluded (e.g., Belczynski et al. 2016a; 2016c)\nocite{bel16a,bel16c} In order to constrain the kinematics of the BBHs, multiple and precise spin measurements from GW detections will be important. Dynamical interactions in GCs can produce highly spinning BHs that are bound with an another BH. If spins are aligned with the binary's orbital angular momentum it points out to the field origin. However, the origin of BBHs with significant spin misalignement would be hard to determine the origin of the binary. 

As pointed out in \citet{Abbott16c}, GW150914 could have been formed either in isolated evolution of low-metallicity binary system or in a dense GC. With the prospects of Observation runs 2 and 3, the advanced LIGO will find more GW sources. The advanced Virgo would be soon up and running to search the GW sky as well. The first GW source catalogue is likely to be dominated by BBHs. GW detections with the advanced LIGO-Virgo detector network will provide an empirical mass distribution for BH populations up to cosmological distances. More detections, in particular for heavy BBHs (with good mass estimates), will allow us to discriminate assumptions on BH mass functions and the plausibility of clusters as BBH factories. In particular, if BHs with mass above pair-instability pulsation supernovae limit ($M_{\rm BH}>~50$ \msun; Belczynski et al.\ 2016\nocite{bel16b,bel16d}) are found in merging binaries, it would be most likely to be produced by mergers between stellar-mass BBHs. The existence of heavier BHs with a few hundred solar mass can only be explained by stellar interactions in dense stellar environments.

GW observations of BBHs will be useful to understand how different BH populations are formed and evolved in different environments. Precise measurements of individual spins and masses will be important to understand underlying properties of BBHs. BBHs. If accessible, electromagnetic wave observations of host galaxies of known BBHs would provide constraints to discriminate different BBH formation and evolution channels.

\section*{Acknowledgements}

This work was partially supported by National Research Foundation (NRF) grants (No.\ NRF-2006-0093852 and NRF-2015R1D1A1A01060201) funded by the Korean government. CK and HM are also grateful for supports by the KISTI (Korea Institute of Science and Technology Information) through collaborative research program in 2016. KB acknowledges support from the NCN grants Sonata Bis 2 (DEC-2012/07/E/ST9/01360), OPUS (2015/19/B/ST9/01099) and OPUS (2015/19/B/ST9/03188).

%%%%%%%%%%%%%%%%%%%%%%%%%%%%%%%%%%%%%%%%%%%%%%%%%%%%%%%%%%%%%%%%%%
%                       REFERENCES
%%%%%%%%%%%%%%%%%%%%%%%%%%%%%%%%%%%%%%%%%%%%%%%%%%%%%%%%%%%%%%%%%%

\end{document}